\documentclass[twocolumn,showpacs,preprintnumbers,amsmath,amssymb]{revtex4}
\usepackage[dvips]{graphicx}% Include figure files
\usepackage{dcolumn}% Align table columns on decimal point
\usepackage{bm}% bold math

\begin{document}

%\preprint{APS/123-QED}

\title{Electrodynamic trapping of spinless neutral atoms with an atom chip}
\author{T. Kishimoto$^{1,\dagger}$, H. Hachisu$^{2}$, J. Fujiki$^{2}$, K. Nagato$^{3}$, M. Yasuda$^{2,\ddagger}$, and H. Katori$^{1,2,*}$}
\affiliation{%
$^1$PRESTO, Japan Science and Technology Agency, Bunkyo-ku, Tokyo 113-8656, Japan\\
$^2$Department of Applied Physics and
$^3$Department of Engineering Synthesis, School of Engineering, 
The University of Tokyo, Bunkyo-ku, Tokyo 113-8656, Japan
}

\date{\today}% 
\begin{abstract}
Three dimensional electrodynamic trapping of neutral atoms has been demonstrated.
By applying time-varying inhomogeneous electric fields with micron-sized electrodes, nearly $10^2$ 
strontium atoms in the $^1S_0$ state have been trapped with a lifetime of 80~ms. 
In order to design the electrodes, we numerically analyzed the electric field and simulated atomic 
trajectories in the trap, which showed reasonable agreement with the experiment.  
\end{abstract}

% PACS, the Physics and Astronomy
                             % Classification Scheme.
%\keywords{Suggested keywords}%Use showkeys class option if keyword
                              %display desired
\maketitle

Coherent manipulation of atoms or ions in the vicinity of solid surfaces has attracted increasing interest 
as a promising tool for quantum information processing (QIP), because of their potential scalability and 
controllability of atoms or ions that work as qubits \cite{Wineland, Cirac, Schmidt-Kaler, Leibfried}.
So far, two approaches, {\it i.e.}, magnetic manipulation of atoms with miniaturized wire traps 
\cite{Fortagh,Haensch} and miniature Paul traps \cite{Wineland} for ions, 
have been demonstrated.
Recent experiments, however, have witnessed that coherence time of these trapped atoms or ions 
was shortened by electro-magnetic interactions caused by thermal magnetic fields 
\cite{B,Harber,Leanhart,Lin} or fluctuating patch potentials \cite{Turchette} appeared on the surface if the distance
between the particle and the surface become smaller than 100~$\mu$m.
To avoid these harmful influences and have a lifetime nearly a second, paramagnetic atoms need 
to be more than tens of microns apart from metal surfaces at room temperature \cite{B,Leanhart,Harber,Lin}.
A reported heating rate of ions \cite{Turchette} indicated stronger coupling of trapped ions 
to surface potentials than that of neutral atoms.

It has been pointed out that the best candidates for long-lived trap are spinless neutral atoms, which weakly 
interact with stray fields via the Stark effect \cite{Henkel,Henkel2}.
Alternatively, material dependence of the trap lifetime has been investigated to reduce thermal magnetic 
field in magnetic atom-chips \cite{Harber,Lin,Sinclair}. 
Electric manipulation of atoms, which allows manipulating spinless neutral atoms in addition to paramagnetic 
atoms and molecules, may open up a new possibility for scalable quantum systems with long coherence time.
In this Letter, we demonstrate three dimensional (3D) electrodynamic trapping of laser cooled Sr atoms 
in the ${}^1S_0$ state with miniature electrodes fabricated on a glass plate.
The very thin electrodes ($\sim 40$~nm) used in the experiment will significantly reduce thermal magnetic fields near metal surfaces, which would be especially profitable in applying this scheme to paramagnetic atoms. 
%%%Furthermore, very low energy consumption of the electric trapping will guarantee dense integration and scalability of the atom circuit.

For an applied electric field ${\bf E}({\bf r})$, the Stark energy is given by 
$U({\bf r})=-\frac{1}{2}\alpha |{\bf E}({\bf r})|^2$.
Since the static dipole polarizability $\alpha$ is positive for atoms in stable states, these atoms can be 
trapped at a local maximum of the electric field strength and behave as a ``high-field seeker". 
However, as the Laplace equation does not allow an electrostatic field to form a maximum in free space, 
3D trapping is not possible for a static electric field alone \cite{Wing}.
In addition owing to a small dipole polarizability, rather high electric fields are required for 
the Stark manipulation of laser-cooled atoms: So far 1D or 2D focusing/trapping experiments 
\cite{Shimizu_2Dtrap,Salomon,Noh} have been demonstrated by applying several to ten kV to electrodes 
with dimensions of a few mm. 
A dynamic stabilization scheme, as employed in RF ion traps, allows electric trapping with higher dimensions. 
Electrodynamic 2D focusing of atoms \cite{Shimizu_2Dtrap} and guiding of molecules \cite{Rempe} were 
demonstrated by using 4 rods with oscillating voltages. 3D trapping by 3 phase electric dipole fields 
\cite{Shimizu_etrap} or by an oscillating hexapole field superimposed on a static homogeneous field 
\cite{Peik} has been proposed. The latter scheme has recently been demonstrated in trapping cold polar molecules 
\cite{Meijer}.

\begin{figure}[b]
\begin{center}
\includegraphics[width=0.9\linewidth]{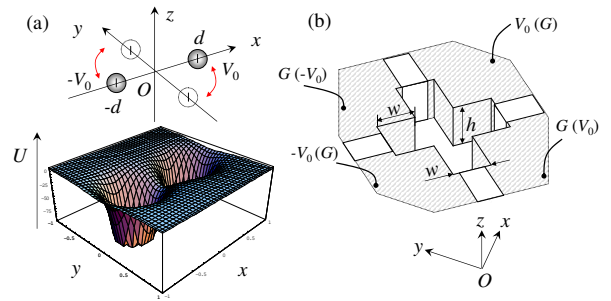}
\caption{
Configuration of electrodes. (a) For analytic calculation, we assume two sphere electrodes applied with 
$\pm V_0$ and located at $\pm d$ on the $x$ axis, which produce Stark potential of $u_x$ as plotted below.
(b) Actual electrodes are fabricated on a glass plate of thickness $h$, with through-hole crossed channels 
of width $w$. The shaded regions are silver-coated to form electrodes, which are set to either $\pm V_0$ or 
ground level (G). Atoms will be trapped at the center of the through-hole.
}
\label{fig1}
\end{center}
\end{figure}

Here we consider 3D electrodynamic trapping with two-phase electric-dipole field, which will allow planar 
geometry that is suitable to be used with atom chips \cite{Katori_e-trap}. 
In order to illustrate the scheme, we first assume two spherical electrodes with radius $b$ placed at $ \pm d$ 
either on the $x$ (or $y$) axis, kept at voltages $\pm V_0$ as shown 
in Fig.~\ref{fig1}(a).
The corresponding Stark energy $u_x$ (or $u_y$) for an atom near the origin is calculated in the lowest order 
of $|x/d|,|y/d|,|z/d|\ll 1$, 
\begin{equation}
\left( {\begin{array}{c}
   {u_x }  \\
   {u_y }  \\
\end{array}} \right) \approx \frac{1}{2}m\omega _0^2 \left( {\begin{array}{r}
   { - \eta x^2  + y^2  + \xi z^2  - d^2 /3}  \\
   {x^2  - \eta y^2  + \xi z^2  - d^2 /3}  \\
\end{array}} \right)
\label{EQ:stark}
\end{equation}
where $m$ is the atomic mass, $\omega_0=(2b V_0/d^3) \sqrt{3\alpha/m}$ is the trapping frequency of an atom 
perpendicular to the dipole axis.
The axial symmetry of electrodes determines $\eta=2$ and $\xi=1$. 
The static dipole polarizability of 
$\alpha=3.08\times10^{-39}\,{\rm J/(V/m)}^2$ \cite{Hyman} for Sr atoms in the $^1S_0$ ground state is used 
in the calculation.
While both of the potentials $u_x$ and $u_y$ provide static confinement along the $z$ axis, they form saddle 
potential in the $xy$-plane. 
The 3D trapping, therefore, can be realized by dynamically stabilizing atomic trajectories in $xy$-plane 
by alternating these two potentials. 
In Eq.~(1) anharmonic terms higher than $r_i^2 r_j^2/d^4$, where $r_i$ and $r_j$ stand for $x,y,z$ coordinate, 
are neglected. However, as discussed later, these terms play crucial role in determining the effective trap volume. 

\begin{figure}
\begin{center}
\includegraphics[width=0.597\linewidth]{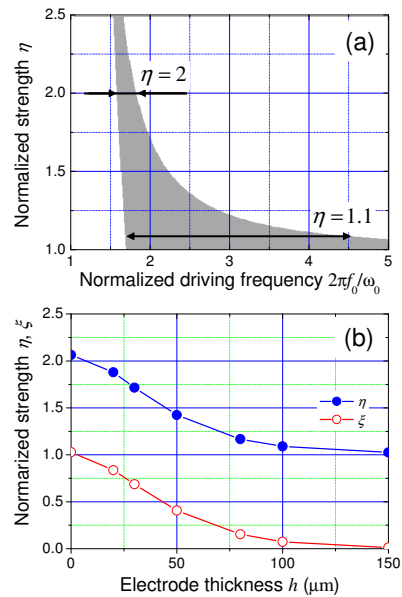}
%\scalebox{0.33}{\includegraphics{Figure1_10.eps}}  %Figure1_2.eps
\caption{
(a) The shaded area shows the stable trapping regime with respect to the driving frequency $f_0$ 
and the relative strength $\eta$ of anti-trapping potential.
The arrows indicates the stability region for $\eta=2$ and $\eta=1.1$. (b) Reduction of the normalized 
strength $\eta$ (filled circles) of anti-trapping and $\xi$ (empty circles) of static trapping as a function 
of the electrode thickness $h$ for the channel width of $w=50\,\mu$m.
}
\label{fig2}
\end{center}
\end{figure}

By switching the charge distribution between $x$ and $y$ axis at a period of $T/2$, the time-dependent 
Stark potential is given by
\begin{eqnarray}
	U({\bf r},t)= \left\{
	\begin{array}{ll}
	u_x({\bf r});\quad nT\leq t<(n+\frac{1}{2})T\\
	u_y({\bf r});\quad (n+\frac{1}{2})T\leq t<(n+1)T
	\end{array}
	\right.,~
	\label{eqn:timedep_pt}
\end{eqnarray}
where $n$ is an integer. 
The time evolution of the position and velocity of an atom subjected to $ U({\bf r},t)$ can be described 
by transfer matrices \cite{Ion_Mathieu,Noh}, whose eigen-values of $|\varepsilon|\leq 1$ guarantee stable 
trapping. 
Harmonic approximation of the Stark potential as given in Eq.~(1) is used to determine the stability 
of the trap with respect to the driving frequency $f_0\equiv 1/T$.
For the electrodes configuration with $\eta=2$ (and $\xi=1$) as discussed above, 
the stability regime is calculated to be $1.56<2 \pi f_0/\omega_0<1.80$ \cite{Katori_e-trap} 
(see Fig.~\ref{fig2}(a)). 
This narrow stability regime for $f_0$, however, made experiments rather challenging.

The stability regime on $f_0$ can be extended by reducing the normalized strength $\eta$ of the anti-trapping 
potential, as shown by a gray area in Fig.~\ref{fig2}(a).
Wider driving frequency range can be obtained for $\eta$ close to unity, which is realized by applying 
line charges instead of point charges.
Figure 1(b) depicts a model for designing actual electrodes, where shaded parts are made 
of conducting material and the other parts insulator.
When voltages are applied to four rectangular parallelepiped electrodes with diagonal separation 
of $\sqrt{2}w$ and thickness of $h$, most of the charges distribute at the ridge of the electrodes. 
Therefore the Stark potential can be approximated by line charges of length $h$.

We numerically analyzed the electric field produced by these electrodes by employing the finite element 
method (FEM).
Figure~\ref{fig2}(b) shows the normalized strength of the anti-trapping potential $\eta$ (filled circles) 
and that of the static potential $\xi$ (empty circles) as a function of the electrode thickness $h$, 
where channel width of $w=50\,\mu$m was assumed.
For thinner electrodes, $\eta\approx 2$ and $\xi\approx 1$ were obtained in agreement with the case 
of two point charges \cite{Katori_e-trap} because the charge distributes at the tip of the electrode.
By increasing the thickness of the electrodes, while the trapping frequency $\sqrt{\xi} \omega_0$ 
along $z$-axis becomes weaker, $\eta$ approaches to unity and wider stability region can be obtained, 
as expected for the two dimensional atom guide \cite{Shimizu_2Dtrap,Rempe}.

\begin{figure}
\includegraphics[width=\linewidth]{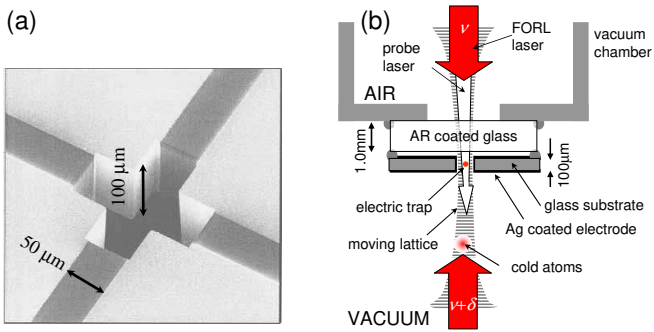}
\caption{(a) A scanning ion microscope (SIM) image of a Stark chip fabricated on a 100-$\mu$m-thick glass 
substrate. 
The dark and bright faces correspond to exposed glass and silver electrodes, respectively. 
(b) A schematic of the experimental setup. Laser-cooled ${}^{88}\mathrm{Sr}$ atoms were transferred 
into an electric trap by a moving lattice. The electric trapping was performed while the moving lattice lasers 
were off.}
\label{fig3}
\end{figure}

Considering these trade-offs, we designed the thickness of the electrodes as $h$=100~$\mu$m for $w=50\,\mu$m, 
which provided $\eta=1.1$ (see Fig.~2(b)). 
By applying voltages of $V_0=200$~V to the electrodes, $\omega_0=2\pi\times2.9$~kHz and $\sqrt{\xi} 
\omega_0=2\pi\times0.89$~kHz along $z$-axis were obtained. 
The latter is strong enough to support atoms against gravity even when $z$-axis directs vertically.
The electrode assembly was made on a 100-$\mu$m-thick fused-silica substrate of 1 inch diameter, 
which was first coated with 250~nm thick silver on both sides.
A cross through-hole and electrode pattern were fabricated by Focused Ion Beam (FIB) process.
Side-walls of the hole were then coated with 40-nm-thick silver to form four electrodes. 
Figure \ref{fig3}(a) shows a scanning ion microscope (SIM) image of the atom chip.

An electrodynamic trapping experiment was performed in four steps, in less than a second: 
(1) Laser cooling and trapping of $^{88}$Sr atoms, (2) transport of atoms into an electric trap, 
(3) electrodynamic trapping, and (4) detection of trapped atoms.
A schematic of an experimental setup is shown in Fig.~\ref{fig3}(b).
We cooled and trapped more than $10^4$ atoms at several $\mu$K in 0.6~s by using magneto-optical trap (MOT) 
on the ${}^1\!S_0-{}^3\!P_1$  spin-forbidden  transition at $\lambda=689$~nm \cite{Katori}.  
Silver surfaces of the electrodes as shown in Fig.~\ref{fig3}(a) were used 
as a mirror to perform a mirror MOT \cite{Haensch}, in which the trapped atom cloud 
was located 1.5~mm below the mirror surface. 
This electrode assembly, or the ``Stark chip" was glued on a 1-mm-thick vacuum view port 
with a clear aperture of 8~mm.
%%% and is anti-reflection coated at around 800 nm. 
The vacuum pressure was typically $1\times 10^{-9}$~Torr during the experiment.

The atoms were then loaded into a one dimensional far-off-resonant optical lattice (FORL) formed 
by a pair of counter-propagating lasers at $\lambda_{L}\approx 810$~nm \cite{Katori1999}. 
These lasers were focused onto the center of the through hole of the Stark chip. The $1/e^2$ waist radius 
was set to 16~$\mu$m so as to have the Rayleigh length of 1~mm to reach atoms in the MOT. 
Laser intensity of 100~mW per beam was chosen to provide a lattice potential with a radial confinement frequency
of $\omega_{\rm r}\approx 2 \pi \times 1.2~{\rm kHz}$ at the chip center, which was close to a
secular frequency of the electric trap as described below and allowed a good mode matching in atom loading into the electric trap.
%%%The depth of the lattice potential was 11~$\mu$K at the center of the MOT.
%%%Typically 40~$\%$ of atoms were transferred from the MOT to the FORL.
By changing the frequency of one of the lattice lasers \cite{Salomon_ML}, we adiabatically transported atoms 
into the atom chip in 6.8~ms. 
The transported atomic cloud inside the electrode gap had a temperature of about 7~$\mu$K and a
$1/e^2$ radius of $r_{\rm atom} \approx 7~\mu{\rm m}$.

Turning off the lattice lasers, the electrodynamic trapping was started by applying $V_0= 200$~V 
onto the diagonal electrodes (see Fig.~1(b)) at a driving frequency of $f_0\approx 6.4$~kHz, 
which gave a secular frequency of $\approx 1.0$~kHz.
After a certain trapping time, the electric trap was switched off. 
Using the moving lattice, we extracted the atoms 0.28~mm below the chip to observe trapped atoms. 
We illuminated them with a 10-$\mu$s-long probe laser resonant to the ${}^1\!S_0-{}^1\!P_1$ transition 
at $\lambda=461$~nm.
The fluorescence was imaged onto an Intensified CCD (ICCD) camera to measure the number of trapped atoms with an
uncertainty of 10~$\%$.

\begin{figure}
\includegraphics[width=0.597\linewidth]{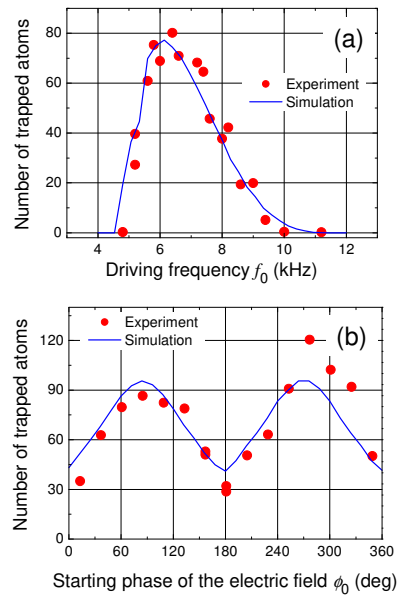}
\caption{ The stability condition for the electro-dynamic trapping. 
Atoms were trapped for 5~ms with driving voltages of $V_0=200$~V. Filled circles and solid lines show 
the experiment and simulation, respectively.
(a) Number of trapped atoms as a function of the driving frequency $f_0$. The simulation took anharmonicity 
of the Stark potential into account.
(b) Number of trapped atoms versus the starting phase $\phi_0$ of the driving electric fields. 
An initial atomic cloud of 7~$\mu$K and  $r_{\rm atom} \approx 7~\mu{\rm m}$ were assumed in the simulation.}
\label{fig4}
\end{figure}

Filled circles in Fig.~\ref{fig4}(a) show the number of trapped atoms as a function 
of the driving frequency $f_0$ for a trapping time of 5~ms. 
We have performed numerical integrations of the equation of motion of an atom subjected 
to the alternating electric fields \cite{Katori_e-trap} that were calculated by the FEM 
for the electrode configuration shown in Fig.~1(b).
Taking the initial atom temperature and its spatial distribution as used in the experiment, 
the calculation (solid line) well reproduced the experiment, where the amplitude (atom number) 
was used as a fitting parameter.

As mentioned earlier the stability of the miniaturized electric trap is crucially affected 
by the anharmonicity of the trapping potential, since the Stark potential provided 
by the dipole electric field contains relatively-large higher-order terms
\cite{FEM}.
These terms limit the effective trap diameter 2$r_{\rm eff}$ 
to be typically one fifth of the electrode separation $\sqrt{2}w$ as determined by numerical simulation \cite{Katori_e-trap}. 
This limited trapping volume, in turn, makes the capture velocity $v_{\rm c}$ 
of the trap be dependent on a starting phase $\phi_0$ of electric trapping field.
Atoms with an outward velocity $v_{\rm c}$ toward a particular direction, say $x$-axis, 
need to be decelerated by the driving field before reaching $r_{\rm eff}$ by the end of the trapping-phase, 
while the similar but $180^{\circ}$ out of phase discussion applies for the atomic motion in $y$-axis 
that is in the anti-trapping phase.
We defined $\phi_0=0$ when the Stark potential $U({\bf r},t)$ was switched at $t=n T$ as given in Eq.~(2), 
i.e., $\phi_0=\frac{t-nT}{T} \times 360^{\circ}$ for the $n$-th driving period, and studied the starting phase $\phi_0$ dependent trap efficiency as shown in Fig.~\ref{fig4}(b). 
The number of trapped atoms (filled circles) was in reasonable agreement 
with a numerical simulation (solid line), where the amplitude (atom number)
was used as an adjustable parameter, indicating that the best loading is realized 
for the trapping field started at $\phi_0=90^\circ$ or 270$^\circ$.
In the experiment, a slight asymmetry in the peak heights was observed. This may be attributed to spatial offset of an initial atomic cloud with respect to the trap center, which was possibly caused by misalignment of the FORL lasers.
We have measured the lifetime of atoms in the electric trap to be 80~ms, which was in reasonable agreement with glancing-collisions-limited lifetime \cite{Bjorkholm1988} assuming the background gas pressure of $\sim 10^{-8}$~Torr in the electrode gap.
Note that similar lifetime was observed for atoms in the FORL, when its trap depth was comparable to that of the electric trap ($\approx 30 \mu$K).

In applying atom-traps to QIP experiments, qubit states should experience the same trapping potential 
so as to minimize decoherence caused by atomic motion \cite{Katori2003,Haroche2004}.
For example, in the case of this Stark trap, $\pm m$ Zeeman substates of the $^3P_2$ metastable state, 
which has a lifetime over 100 s \cite{Yasuda2004}, can be used as a qubit state that experience 
the same Stark shift.
Although the coherent evolution of these states may be disturbed by the thermal magnetic fields 
that appeared on the electrodes surfaces, a very thin electrode (40~nm) demonstrated here may significantly 
reduce thermal magnetic fields that cause spin flips \cite{Henkel,Henkel2}, since the Johnson noise 
induced currents decrease as electrode's thickness \cite{Lin}.
Furthermore, since the operation of the electrodynamic trap relies on switching of electric fields, 
it is free from ohmic dissipation and allows dense integration of traps.
Array of electrodes, four of which activated in turn so as to adiabatically transfer atoms, 
may constitute atom wave guide that is reminiscent of the quantum CCD \cite{Kielpinski}.

In conclusion, we have investigated the design of electrodes with a help of numerical simulation 
and demonstrated an electrodynamic trapping of spinless neutral Sr atoms
with micron-sized structures.
By reducing the electrode size to a few $\mu$m, these atom traps can be driven by a few volts 
\cite{Katori_e-trap}, which will make electric atom traps compatible with electronic logic circuits, 
offering an interface between atom manipulation and electronics.

The authors would like to thank A. Yamauchi, and M. Tange for their technical support.


\begin{thebibliography}{1}
\bibitem{FEM}
		The FEM analysis indicated that the contribution of the higher order terms is on the order of $10^{-2}$ of the total potential energy for an atom at $r_{\rm eff}$=8~$\mu$m from the center.
\end{thebibliography}
\end{document}